\documentclass{article}
\usepackage{graphicx}
\usepackage{amsmath,amssymb}
\usepackage{graphics,color,array,calc,rotating,epsfig,psfrag}
\usepackage{hyperref}
\usepackage{tikz}
\usetikzlibrary{positioning}
\numberwithin{equation}{section}
\usepackage{cite}
\usepackage{bm}
\usepackage{dcolumn}
\usepackage{float}
\usepackage{subfigure}
\usepackage{tikz-cd}
\usepackage{amsfonts}
\usepackage[mathscr]{eucal}
\usepackage{helvet}
\def\be{\begin{equation}} \def\ee{\end{equation}}
\def\bea{\begin{eqnarray}} \def\eea{\end{eqnarray}}

\newcommand{\labell}[1]{\label{#1}} 
\newcommand{\reef}[1]{(\ref{#1})}

\newcommand{\RN}[1]{%
  \textup{\uppercase\expandafter{\romannumeral#1}}%
}
\begin{document}
\baselineskip 18pt%
\begin{titlepage}
\vspace*{1mm}%
\hfill%
\vspace*{15mm}%
\hfill
\vbox{
    \halign{#\hfil         \cr
          } 
      }  
\vspace*{20mm}
\begin{center}
{\Large {\bf \boldmath  Manifestly $SL(2,R)$ Duality-Symmetric Forms in ModMax Theory}}
 \end{center}
\vspace*{5mm}
\begin{center}
{H. Babaei-Aghbolagh$^{a}$, Komeil Babaei Velni$^{b}$,
	 Davood Mahdavian Yekta$^{c}$ and H. Mohammadzadeh$^{a}$
 }\\
\vspace*{0.2cm}
{\it
$^{a}$Department of Physics, University of Mohaghegh Ardabili,
P.O. Box 179, Ardabil, Iran\\
$^{b}$Department of Physics, University of Guilan, P.O. Box 41335-1914, Rasht, Iran\\
$^{c}$Department of Physics, Hakim Sabzevari University, P.O. Box 397, Sabzevar, Iran\\
}

 \vspace*{0.5cm}
{E-mails: {\tt h.babaei@uma.ac.ir,  babaeivelni@guilan.ac.ir, d.mahdavian@hsu.ac.ir, mohammadzadeh@uma.ac.ir,
}}
\vspace{1cm}
\end{center}

\begin{abstract}
In this paper, we will investigate a manifestly $SL(2,R)$-invariant structure for the energy-momentum tensor of ModMax theory as a nonlinear modification of Maxwell electrodynamics which includes conformal invariance as well. In the context of this theory, we show that the energy-momentum tensor of the generalized Born-Infeld theory can also be written in the same invariant form. We will find manifestly self-dual invariant actions corresponding to the invariant couplings $\lambda$ and $\gamma$ in these theories. It can be shown that the resultant actions correspond to the irrelevant and marginal $T\bar{T}$-like deformations, respectively.
\end{abstract}

\end{titlepage}

\section{Introduction}\label{1}
It has been found different classes of non-linear electrodynamic (NED) theories as candidates to the extension of Maxwell electrodynamics in the literature \cite{Plebanski,Boillat,Birula1,Birula2}. Also, one can find some recent developments in this area of research in Refs. \cite{Aschieri:2008ns,Dehghani:2021fwb,Sorokin:2021tge,Paixao:2205jsw}. Recently, it has been discovered a nonlinear extension of Maxwell's theory, which  not only is invariant under electromagnetic duality transformations, but also is a conformally invariant theory. This electromagnetic modification is known as the ModMax theory \cite{Bandos:2020jsw} and is described by the following Lagrangian density
\begin{eqnarray}\label{LMM}
{\cal L}_{MM}=\cosh(\gamma) \mathcal{S}+\sinh(\gamma) \sqrt{ \mathcal{S}^2+\mathcal{P}^2},
\end{eqnarray}
where $ \mathcal{S}=-\frac{1}{4}F_{\mu\nu}F^{\mu\nu}$ and $\,\,\mathcal{P}=-\frac{1}{4}F_{\mu\nu}\tilde F^{\mu\nu}$ are two Lorentz invariant variables, which $\tilde{F}^{\mu \nu}= \frac{1}{2} \varepsilon^{\mu \nu\alpha \beta} F_{\alpha \beta}$ is the Hodge dual of the electromagnetic field strength $F_{\mu \nu}$. As is obvious from (1.1), the ModMax theory comprises a positive constant $\gamma$ so that for $\gamma=0$, it reduces to the Maxwell Lagrangian \cite{Kosyakov:2020wxv}. The generalization of the ModMax theory in the context of conformal invariant p-form gauge theories was studied in Ref. \cite{Bandos:2020hgy}.

It has been shown \cite{Bandos:2020jsw,Bandos:2020hgy} that the ModMax and the Born-Infeld theories can be combined into a duality-symmetric generalized Born-Infeld (GBI) electrodynamics as follows
\begin{equation}\label{GBI}
{\cal L}_{BI \gamma} = \frac{1}{\lambda} \Bigg[ 1 -  \sqrt{1 -  \lambda \Bigl( 2 {\cal L}_{MM}+\lambda \mathcal{P}^2 \Bigr)} \Bigg],
\end{equation}
where in the weak field limit $\lambda\to 0$, the Lagrangian \reef{GBI} reduces to the ModMax theory denoted by \reef{LMM}. More studies on the ModMax theory, such as the supersymmetric actions, the nonlinear generalizations, and solutions of field equations can be found in Refs.~\cite{Bandos:2021rqy,Kuzenko:2021cvx,Avetisyan:2021heg,Kruglov:2021bhs,BallonBordo:2020jtw,Nastase:2021uvc,Mkrtchyan:2205uvc,Barrientos:2022bzm}.  Also, an alternative form of the ModMax Lagrangian  in terms of axion-dilaton-like auxiliary scalar fields coupling has been proposed in Ref. \cite{Lechner:2022qhb}. ModMax theory has been more carefully studied and expanded in recent works \cite{Barrientos:2022bzm,Bokulic:2022cyk,Banerjee:202207bzm,Pantig20220bzm,Neves20220bzm}.

As well known, the Maxwell's field equations are invariant under the electric-magnetic $SO(2)$ duality transformations, however, the enhancement of this symmetry to the level of an action is another problematic task. This fact becomes even more complicated for theories featuring nonlinear interactions of the Maxwell fields. There are three approaches to construct theories of self-interacting Maxwell fields incorporating an electromagnetic $SO(2)$ duality invariance; the Gaillard and Zumino \cite{Gaillard:1981rj, Gaillard:1997rt} approach which has been further developed by Gibbons and Rasheed \cite{Gibbons:1995ap}, the non-covariant first order Hamiltonian approach based on the work by Henneaux and Teitelboim \cite{HT} and further followed by Deser, Gomberoff, Henneaux and Teitelboim \cite{Deser:1997mz,Deser:1997se}, and the third one is the PST method proposed by Pasti, Sorokin and Tonin \cite{Pasti:1995tn,Pasti:1995us,Pasti:1996vs,Pasti:1997gx}.

In the first systematic method to construct duality invariant theories, we have to define a doublet $\left(\tilde{F},G\right)$,  where $G_{\mu\nu}=-2\, \frac{\partial { \cal L} (\mathcal{S},\mathcal{P})}{\partial F^{\mu\nu}}$ is an antisymmetric tensor, to show that the equations of motion are invariant under $SO(2)$ rotations. This requirement imposes a non-linear differential equation; $G\tilde{G}-F\tilde{F}=0$, the so-called self-dual condition. In the second approach, $SO(2)$ duality is an exact and explicit symmetry of the Lagrangian of the theory. Such a NED theory has a doublet of electric and magnetic fields $(A_1,A_2)$ and the $SO(2)$ duality is an ordinary 2d rotation on these doublets. Finally in the PST approach, if one introduces an auxiliary scalar field and a compensating gauge symmetry, it is possible to formulate a manifestly covariant action with the correct degrees of freedom.

We will follow the Gaillard-Zumino approach and present the self-dual invariant actions for the ModMax and GBI (and axion-dilaton-GBI) theories which gives the correct equations of motion when one imposes the duality invariant condition as an extra constraint. As the main purpose of our study in this letter, we find a manifestly $SL(2,R)$ invariant structure for the energy-momentum tensors of the ModMax and also the GBI theories. In this respect, although the corresponding Lagrangians are not invariant under $SO(2)$ duality-symmetry, their derivatives with respect to the invariant parameters are. The invariant parameter could be a coupling constant or an external background field, such as the gravitational field, which does not change under duality rotations\cite{Gaillard:1981rj}.
In this approach if a theory satisfies the self-dual condition, then the physical objects of the theory such as equations of motion\cite{Green:1996qg}, energy-momentum tensor\cite{Gibbons:1995ap,BabaeiVelni:2016qea} and scattering amplitudes\cite{Babaei-Aghbolagh:2013hia,Garousi:2017fbe,BabaeiVelni:2019ptj} are (S-dual) invariant. We are also interested in finding corresponding self-dual invariant theory in the form that are manifestly $SL(2,R)$ invariant.
The self-dual invariant Lagrangian is given by the derivative of the original Lagrangian with respect to the constant parameters $\gamma$ and $\lambda$
 \cite{Aschieri:2013nda}.

The structure of this paper is organized as follows: in Sec. \ref{2}, we enhance $SO(2)$ duality in the ModMax  as well as the GBI theories to the non-compact $SL(2, R)$  group by coupling the electromagnetic field to the dilaton and axion fields. In Sec. \ref{secLinv}, we find the manifestly self-dual invariant actions for a general non-linear GBI theories of electrodynamics. We show that  the expansion of a self-dual invariant Lagrangian  in NED theories  is compatible with  the deformed Lagrangian constructed by both irrelevant and marginal  $T \bar{T}$ operators in NED theories. Finally, the Sec.~\ref{4} is devoted to giving a brief summary of results and identifies some directions for future researches.
\section{$SL(2,R)$ symmetry in axion-dilaton ModMax Theory}\label{2}

The equations of motion of free Maxwell theory
are $\partial_\nu F^{\mu \nu} =0$  and $\partial_\nu \tilde{F}^{\mu \nu} =0$.
These equations  obviously  are transformed into each other under the rotation of $ (F^{\mu \nu},\tilde{F}^{\mu \nu}) \to (\tilde{F}^{\mu \nu},-F^{\mu \nu})$. This duality, which is known as the electric-magnetic  duality, is a special case of the $SO(2)$ duality-symmetry. In this section we are going to study the $SO(2)$ symmetry  for the  energy-momentum tensor of the ModMax theory and its extension to $SL(2,R)$ symmetry  by coupling the electromagnetic field to some axion and dilaton fields. Consequently, we find the $SL(2,R)$ duality-symmetric structures for the energy-momentum tensor of the ModMax and  GBI theories.
\subsection{Electric-magnetic duality in the ModMax theory }
According to \cite{Gibbons:1995cv}, a general electric-magnetic duality rotation through an angle $\alpha$ is given by
\begin{equation}
\left\{
\begin{array}{rcl}
G_{\mu\nu} & \!\!\!\rightarrow \!\!\! & \cos\alpha ~G_{\mu\nu} + \sin\alpha ~\tilde{F} \\
& & \\
\tilde{F}_{\mu\nu} & \!\!\! \rightarrow \!\!\! & \cos\alpha ~\tilde{F}_{\mu\nu}-\sin\alpha ~
G_{\mu\nu}.
\end{array}
\right.
\label{Rot}
\end{equation}
This rotational symmetry is called the $SO(2)$-duality symmetry. Assume a finite $SO(2)$-duality rotation of the angle $\pi/2$, then we have the following relations for the fields and invariant parameters
\begin{equation}\label{GF}
(G^{\mu \nu},\tilde{F}^{\mu \nu}) \to (\tilde{F}^{\mu \nu},-G^{\mu \nu})\,,
\,\,\,\,\,\,\,\,\,\,\,(g_{\mu \nu},\gamma,\lambda) \to (g_{\mu \nu},\gamma,\lambda)\,.
\end{equation}
 One can find the $G$-tensor for the ModMax theory \reef{LMM} as follows
\begin{equation} \label{G}
 G_{\mu\nu}=-2\, \frac{\partial {\cal L}_{MM}}{\partial F^{\mu\nu}}=\cosh(\gamma) F_{\mu\nu} +\sinh(\gamma) \frac{ F_{\mu\nu}\mathcal{S}+ \tilde{F}_{\mu\nu} \mathcal{P}}{ \sqrt{  \mathcal{S}^2+\mathcal{P}^2 } }.
\end{equation}

It has been found in Ref. \cite{Gibbons:1995ap} that the energy-momentum tensor of a NED theory can be written as
\begin{equation}
\label{EMT1}
T_{\mu\nu}=g_{\mu\nu}\, {\cal L} (\mathcal{S},\mathcal{P})+{F_{\mu}}^ \rho G_{\nu \rho},
\end{equation}
therefore, by considering the above relation, the energy-momentum tensor of the ModMax theory \reef{LMM} can be found as
\begin{eqnarray}\label{Tmunu}
T_{\mu\nu}= T^{Max}_{\mu\nu} \bigg(\cosh(\gamma) + \frac{ \mathcal{S}}{ \sqrt{ \mathcal{S}^2+\mathcal{P}^2}} \sinh(\gamma)\bigg),
\end{eqnarray}
where $T^{Max}_{\mu\nu}=F_{\mu \rho}{F_{\nu}}^ \rho+ g_{\mu\nu} \mathcal{S}$ is the energy-momentum tensor of the Maxwell theory.
Now, in order to study the behavior of the energy-momentum tensor \reef{Tmunu} under the nonlinear duality transformations \reef{GF}, we define the following symmetric structure that is invariant under these transformations
\begin{eqnarray}\label{Nmunu}
{\cal N}_{\mu\nu}=G_{\mu}{}^{\alpha} G_{\alpha \nu}+\tilde{F}_{\mu}{}^{\alpha} \tilde{F}_{\alpha\nu}.
\end{eqnarray}

Also substituting the ModMax $G$-tensor \reef{G} in Eq.~\reef{Nmunu}, one can find the symmetric invariant structure corresponding to the ModMax theory
\begin{eqnarray}\label{NN}
{\cal N}_{\mu\nu} &=& -2 \cosh(\gamma) \Bigl(\cosh(\gamma) + \frac{\mathcal{S}}{ \sqrt{ \mathcal{S}^2+\mathcal{P}^2}} \sinh(\gamma)\Bigr) T^{Max}_{\mu\nu}\nonumber\\
&& + 2 \sinh(\gamma) \Bigl(\cosh(\gamma)  \sqrt{ \mathcal{S}^2+\mathcal{P}^2} +\sinh(\gamma) \mathcal{S}\Bigr) g_{\mu\nu}.
\end{eqnarray}
Considering the energy-momentum tensor \reef{Tmunu} and the invariant structure \reef{NN}, we obtain a manifestly invariant $T_{\mu\nu}$ of the form
\begin{equation}\label{manifest}
T_{\mu\nu}=- \frac{1}{2 \cosh( \gamma)} \Big[ {\cal N}_{\mu\nu} -\frac{1}{4}  {\cal N}_{\rho}{}^{\rho}  { g}_{\mu\nu} \Big],
\end{equation}
where $ {\cal N}_{\rho}{}^{\rho}=8\sinh(\gamma) \Bigl(\cosh(\gamma)  \sqrt{ \mathcal{S}^2+\mathcal{P}^2} + \sinh(\gamma) \mathcal{S}\Bigr)$ is the trace of symmetric structure ${\cal N}_{\mu\nu}$. Note also that ${\cal N}_{\rho}{}^{\rho}$ vanishes at $\gamma=0$ and we achieve the standard Maxwell energy-momentum tensor.
\subsection{ModMax theory  coupled to an axion-dilaton field }\label{MMAD}
Consider the following nonlinear $SL(2,R)$ transformation
\begin{eqnarray}
\label{sl2r}
\tau \rightarrow \frac{p\tau+q}{r\tau+s},\,\,\,\,\,\,\,\,\,\,\,\,\,\,\,\,\Lambda=\left(
\begin{array}{cc}
p&q  \\
r&s
\end{array}
\right)\in\, {SL(2,R)},
\end{eqnarray}
where  $ \tau = C_0 + i e ^ {- \phi_0}$ is a complex axion-dilaton field.
Now, for a Lagrangian in which the general NED theory is coupled with the axion field so that the equations of motion are $SL(2,R)$ invariant, one should  add  the term $\frac{1}{4} C_0  F_{\mu \nu} \tilde{F}^{\mu \nu}$ to the primary NED Lagrangian. On the other hand, the contribution of dilatonic field is given by the  field redefinition $\bar{F}_{\mu \nu} \to  e^{-\frac{\phi_0}{2}}F_{\mu \nu}$ \cite{Gaillard:1997rt}. Thus, the final theory appears an axion-dilatonic NED theory denoted by Lagrangian density $\hat{{\cal L}}(\tau , F )=\bar{{\cal L}}(e^{-\frac{\phi_0}{2}}F)+\frac{1}{4} C_0  F_{\mu \nu} \tilde{F}^{\mu \nu}$. In this regard, the $SO(2)$ duality group in Eq. \reef{GF} enhances  to the $SL(2,R)$ symmetry group.  Due to this consideration, one can find the ModMax theory coupled to axion-dilaton field as
\begin{eqnarray}\label{hatLMM}
\hat{{\cal L}}_{MM}(\tau , F )=\bar{{\cal L}}_{MM}(e^{-\frac{\phi_0}{2}}F )- C_0 \mathcal{P}=\cosh(\gamma) e^{-\phi_0} \mathcal{S} +\sinh (\gamma)\sqrt{e^{-2\phi_0}( \mathcal{S}^2+\mathcal{P}^2)}- C_0 \mathcal{P}.
\end{eqnarray}

It is clear that the above theory reduces to $\hat{{\cal L}}_{Max}(\tau , F )$ in the limit of $\gamma \to 0$. So, the energy-momentum tensor obtained from Lagrangian \reef{hatLMM}, only differs up to a dilaton field factor from the energy-momentum tensor of the pure ModMax theory . This  means that $\hat{T}_{\mu \nu}=  e^{-\phi_0} T_{\mu \nu}$  where the tensor $T_{\mu \nu}$  is given by Eq. \reef{Tmunu}. We can also find  the $G$-tensor corresponding to the axion-dilaton ModMax theory \reef{hatLMM} from $\hat{G}_{\mu\nu}=-2\, \frac{\partial \hat{{\cal L}}_{MM}(\tau , F )}{\partial F^{\mu\nu}}$. Thus, we have the  antisymmetric tensor  $\hat{G}_{\mu\nu}=\bar{G}_{\mu\nu}-C_0\tilde{F}_{\mu\nu}$ where  $\bar{G}_{\mu\nu}=-2\frac{\partial \bar{{\cal L}}_{MM}}{\partial F^{\mu\nu}}= e^{-\phi_0} G_{\mu \nu}$ and $G_{\mu \nu}$ is given in Eq. \reef{G}.

 According to \cite{Gibbons:1995ap}, for any NED theory one can extend the duality transformations \reef{GF} to the corresponding $SL(2,R)$ duality group. In the case of ModMax theory, the corresponding $SL(2,R)$ duality transformations is described by
\begin{equation}\label{SL1}
{\cal F}_{\mu\nu}\equiv\left(
\begin{array}{c}
\tilde{F}_{\mu\nu} \cr  \
\bar{G}_{\mu\nu}-C_0\tilde{F}_{\mu\nu}
\end{array}
\right) \rightarrow  (\Lambda^{-1})^T \left(
\begin{array}{c}
\tilde{F}_{\mu\nu}\cr  \
\bar{G}_{\mu\nu}-C_0\tilde{F}_{\mu\nu}
\end{array}
\right)\,,
\end{equation}
and\footnote{Note that the matrix ${\cal M}_0$ here is the inverse of the matrix ${\cal M}$ in \cite{Gibbons:1995ap}.}
\begin{equation}\label{SL2}
{\cal M}_0=e^{\phi_0}\left(\begin{array}{cc}
|\tau|^2 & C_0\\
C_0 & 1
\end{array}\right) ,\,\,\,\,\,\, {\cal M}_0\longrightarrow \Lambda {\cal M}_0\Lambda ^T\,.\end{equation}
Using the transformations \reef{SL1} and \reef{SL2}, we can find a symmetric structure in the form of $\hat{\cal N}_{\mu \nu}=({\cal F}^T)_\mu{}^\alpha {\cal M}_0 {\cal F}_{\alpha \nu }$ which is manifestly invariant under the $SL(2,R)$ duality transformations. That is,
\begin{eqnarray}\label{NAD}
\hat{\cal N}_{\mu \nu}&=&e^{-\phi_0}\,\tilde{F}_\mu{}^\alpha\,\tilde{F}_{\alpha \nu }+ e^{\phi_0}\, \bar{G}_{\mu}{}^\alpha \, \bar{G}_{\alpha \nu }\nonumber\\
&=& e^{-\phi_0} {\cal N}_{\mu\nu}\,,
\end{eqnarray}
where we have used $\bar{G}_{\mu\nu}= e^{-\phi_0} G_{\mu \nu}$ and Eq. \reef{Nmunu} to obtain the second line.
On the other hand, the energy-momentum tensor corresponding to Eq. \reef{hatLMM} is given as follows
\begin{eqnarray}\label{hatTmunui}
\hat{T}_{\mu \nu}=- \frac{1}{2 \cosh( \gamma)} \Big[ {\hat{{\cal N}}}_{\mu\nu} -\frac{1}{4}  {{\hat{{\cal N}}}_{\rho}}^{\rho}  { g}_{\mu\nu} \Big],
\end{eqnarray}
which is a manifestly  $SL(2,R)$ invariant structure in the axion-dilaton ModMax theory.
\subsection{$SL(2, R)$  symmetry in the GBI theory}

Following the discussion in the previous section, one can construct a GBI theory with $SL(2, R)$  symmetry group by adding an axion-dilaton field to $BI \gamma$ theory given by Eq. \reef{GBI}, i,e, as $\hat{{\cal L}}_{BI \gamma}(\tau , F )=\bar{{\cal L}}_{BI \gamma}(e^{-\frac{\phi_0}{2}}F )- C_0 \mathcal{P}$. Due to this identification we obtain
\begin{equation}\label{GBIAD}
\hat{{\cal L}}_{BI \gamma}(\tau, F)= \frac{1}{\lambda} \Bigg[ 1 -  \sqrt{1 -  \lambda \Bigl( 2 e^{-\phi_0} \Big(\cosh(\gamma)  \mathcal{S} +\sinh (\gamma)\sqrt{( \mathcal{S}^2+\mathcal{P}^2)}\Big)+\lambda e^{-2\phi_0} \mathcal{P}^2 \Bigr)} \Bigg]- C_0 \mathcal{P}.
\end{equation}

Similar to what we have a priori done for the ModMax theory, we are interested in the $SL(2,R)$ invariant structure for the energy-momentum of GBI theory in \reef{GBIAD}.
Therefore, using the standard definition of the energy-momentum tensor and after a straightforward calculation, we obtain
\begin{eqnarray}\label{Tbig}
\hat{T}_{\mu \nu}&=&e^{-\phi_0}\frac{F_{\mu \alpha} F_{\nu}{}^{\alpha} \bigl(\cosh(\gamma) \sqrt{ \mathcal{S}^2+\mathcal{P}^2} + \sinh(\gamma) \mathcal{S}\bigr)}{ x\sqrt{ \mathcal{S}^2+\mathcal{P}^2}}  \nonumber\\
&+& e^{-\phi_0} \frac{\bigl(2 \lambda \cosh(\gamma)\mathcal{S} \sqrt{ \mathcal{S}^2+\mathcal{P}^2}+\lambda \sinh(\gamma) (\mathcal{P}^2 + 2  \mathcal{S}^2)+ e^{\phi_0} ( x-1)\,\sqrt{ \mathcal{S}^2+\mathcal{P}^2} \bigr) \mathit{g}_{\mu \nu}}{\lambda  x \sqrt{ \mathcal{S}^2+\mathcal{P}^2}},
\end{eqnarray}
where, for simplicity, we use the shorthand $x=\sqrt{1 -  \lambda \Bigl( 2 e^{-\phi_0} \Big(\cosh(\gamma)  \mathcal{S} +\sinh (\gamma)\sqrt{( \mathcal{S}^2+\mathcal{P}^2)}\Big)+\lambda e^{-2\phi_0} \mathcal{P}^2 \Bigr)}$.
The $\lambda$-expansion of the above energy momentum tensor up to order $\lambda$ is as follows
\begin{eqnarray}\label{Tlanda}
\hat{T}_{\mu \nu}&=& e^{-\phi_0}\biggl(\cosh(\gamma) + \frac{ \sinh(\gamma) \mathcal{S}}{\sqrt{ \mathcal{S}^2+\mathcal{P}^2}}\biggr) T^{Max}_{\mu \nu} + e^{-2\phi_0} \lambda \biggl[ \Bigl(  \cosh(2 \gamma)\mathcal{S}+\frac{ \sinh(2 \gamma) (\mathcal{P}^2 + 2  \mathcal{S}^2)}{2 \sqrt{ \mathcal{S}^2+\mathcal{P}^2}}\Bigr) T^{Max}_{\mu \nu} \nonumber\\
&-&  \frac{1}{2} \Bigl( \cosh^2(\gamma) \mathcal{P}^2 + \mathcal{S} \bigl(\cosh(2 \gamma)\mathcal{S} - \sinh(2 \gamma) \sqrt{ \mathcal{S}^2+\mathcal{P}^2} \bigr)\Bigr) \mathit{g}_{\mu \nu}\biggr]+{\cal O} ( \lambda^2)+\ldots\, .
\end{eqnarray}
 The leading term in the expansion of \reef{Tlanda} is the energy-momentum tensor of the axion-dilaton ModMax theory that we found in terms of $SL(2,R)$ invariant structure in Eq. \reef{hatTmunui}.
In fact, if the Eq.~\reef{Tlanda} is to be invariant under the $SL(2,R)$ transformations, one should find it invariant at any order of $\lambda$ as well.

 In comparison to the axion-dilaton ModMax theory, the antisymmetric $G$-tensor of generalized theory in Eq. \reef{GBIAD} could be found from   $\hat{G}_{\mu\nu}=-2\, \frac{\partial \hat{{\cal L}}_{BI \gamma}(\tau, F)}{\partial F^{\mu\nu}}$. However, one can write the result as $\hat{G}_{\mu\nu}=\bar{G}_{\mu\nu}-C_0\tilde{F}_{\mu\nu}$, where
\begin{eqnarray}\label{GGne}\bar{G}_{\mu\nu}=\frac{e^{-\phi_0}\sinh(\gamma) (F_{\mu \nu}\mathcal{S} +\mathcal{P} \tilde{F}_{\mu \nu})}{\sqrt{ \mathcal{S}^2+\mathcal{P}^2} x} + \frac{ e^{-\phi_0} \cosh(\gamma) F_{\mu \nu} + e^{-2\phi_0} \lambda \mathcal{P} \tilde{F}_{\mu \nu}}{x}\, .
\end{eqnarray}
The $\lambda$-expansion of the above tensor up to order $\lambda$ is as follows
\begin{eqnarray}\label{Gbar}
\bar{G}_{\mu \nu}&=&e^{-\phi_0}\bigg[\cosh(\gamma) F_{\mu \nu} +  \frac{\sinh(\gamma) (F_{\mu \nu}\mathcal{S} + \tilde{F}_{\mu \nu} \mathcal{P})}{ \sqrt{ \mathcal{S}^2+\mathcal{P}^2}} \bigg]  +e^{-2\phi_0} \lambda \Bigl[ \cosh^2(\gamma) \tilde{F}_{\mu \nu} \mathcal{P}+ \frac{\cosh(\gamma)  \sinh(\gamma) \tilde{F}_{\mu \nu} \mathcal{P}\mathcal{S}}{ \sqrt{ \mathcal{S}^2+\mathcal{P}^2}}\nonumber  \\
&+&\frac{F_{\mu \nu} \bigl(2 \cosh(2 \gamma)\mathcal{S}  \sqrt{ \mathcal{S}^2+\mathcal{P}^2} +  \sinh(2 \gamma) (\mathcal{P}^2 + 2  \mathcal{S}^2)\bigr)}{2 \sqrt{ \mathcal{S}^2+\mathcal{P}^2}}\Bigr]+{\cal O} ( \lambda^2)+\ldots\,.
\end{eqnarray}

Substituting $\bar{G}_{\mu\nu}$ from \reef{Gbar} in Eq. \reef{NAD}, we find  the $SL(2,R)$ invariant structure corresponding to  \reef{GBIAD} up to order $\lambda$, i.e. ,
\begin{eqnarray}
\hat{\cal N}_{\mu\nu} &\!\!\!=\!\!\!&-2 e^{-\phi_0}\bigg[\cosh(\gamma) \Bigl(\cosh(\gamma) + \frac{\mathcal{S} \sinh(\gamma)}{ \sqrt{ \mathcal{S}^2+\mathcal{P}^2}}\Bigr) T^{Max}_{\mu\nu}- \sinh(\gamma) \Bigl(\cosh(\gamma)  \sqrt{ \mathcal{S}^2+\mathcal{P}^2} + \sinh(\gamma) \mathcal{S}\Bigr) g_{\mu\nu}\bigg]\nonumber \\
&\!\!\!+\!\!\!&-2  \lambda \, e^{-2\phi_0} \biggl[\Bigl(\cosh(3 \gamma)\mathcal{S}+ \frac{2 \cosh^2(\gamma) \sinh(\gamma) \mathcal{P}^2 }{ \sqrt{ \mathcal{S}^2+\mathcal{P}^2}} + \frac{ \sinh(3 \gamma)  \mathcal{S}^2}{ \sqrt{ \mathcal{S}^2+\mathcal{P}^2}}\Bigr) T^{Max}_{\mu\nu} \\
&\!\!\!-\!\!\!&\frac12\bigl(\cosh(\gamma) \mathcal{P}^2 + \cosh(3 \gamma) (\mathcal{P}^2 + 2  \mathcal{S}^2) + 2 \sinh(3 \gamma) \mathcal{S} \sqrt{ \mathcal{S}^2+\mathcal{P}^2}\bigr) g_{\mu\nu}\biggr]  +{\cal O} ( \lambda^2)+\ldots\, .\nonumber
\end{eqnarray}

Now, by considering this invariant structure and after performing some lengthy calculations we are able to recast the energy-momentum tensor of the axion-dilaton $BI\gamma$ theory in Eq. \reef{Tbig} in a manifestly $SL(2,R)$ invariant form. Therefore, we have
\begin{eqnarray}\labell{hatT}
\hat{T}_{\mu\nu}=- \frac{1}{2 \cosh( \gamma)} \Big( \hat{\cal N}_{\mu\nu} -\frac{1}{4}  \hat{\cal N}_{\rho}{}^{\rho}  { g}_{\mu\nu} \Big)
+\lambda \biggl[ a_0(\gamma)\, \hat{\cal N}_{\rho}{}^{\rho}\, \hat{\cal N}_{\mu \nu} +b_0(\gamma)\, \hat{\cal N}_{\mu}{}^{\rho}\, \hat{\cal N}_{\nu \rho} \biggr]+\ldots,
\end{eqnarray}
where the coefficients $a_0(\gamma)$ and $b_0(\gamma)$ are given by
\begin{eqnarray}a_0(\gamma)= -  \frac{1}{16 \cosh^2(\gamma)} +\frac{3}{16}, \,\,\,\,\,\,\,\,\,\,b_0(\gamma)= \frac{1}{4\cosh^2(\gamma)}- \frac{3}{8}\,. \nonumber
\end{eqnarray}
 Note that in Eq. \reef{hatT} we have also used the  identities $F_{\mu}{}^{\alpha} \tilde{F}_{\nu \alpha}=- g_{\mu \nu}  \mathcal{P}$ and $ F_{\mu}{}^{\alpha} F_{\alpha}{}^{\gamma} F_{ \gamma \beta} F^{\beta}{}_{\nu}= 2 F_{\mu}{}^{\alpha} F_{ \alpha \nu}\mathcal{S}+ \mathit{g}_{\mu \nu} \mathcal{P}^2$ that hold for any two-form $F$ . The explicit forms of the invariant structures that appear in Eq. \reef{hatT} have mentioned in the appendix.
\section{Manifestly self-dual invariant actions}\label{secLinv}
As mentioned before, the Lagrangian of the Maxwell and the ModMax theories are not invariant under the $SL(2,R)$ transformation, however, the derivatives of them with respect to an invariant parameter are invariant.
In this paper we deal with two coupling constants $\lambda$ and $\gamma$ and also the gravitational field. The energy-momentum tensor that obtains from the variation of the corresponding Lagrangian with respect to the gravitational field would be invariant under this duality\cite{Gaillard:1981rj,Gaillard:1997rt}. The non-zero contribution of the invariant action in electrodynamic theories comes from the variation of original theory with respect to the electromagnetic field strength \cite{Carrasco:2011jv}. It has been shown in \cite{Aschieri:2013nda} that the derivative of the original Lagrangian $\cal L$ with respect to the coupling constant $c$ in the form $c \frac{\partial{ \cal L}}{\partial c}$ produces the corresponding invariant theory $L^{c}_{inv}$. However, for a theory with a typical dimensionless coupling constant such as the ModMax theory, the corresponding derivative can produce an invariant theory.

Due to this fact and according to \reef{GBI}, consider the invariant Lagrangians $L^{\lambda}_{inv}$ and $L^{\gamma}_{inv}$ corresponding to the coupling constants $\lambda$ and $\gamma$, respectively. Obviously, the Lagrangians  of the Maxwell theory as well as the ModMax theory  are independent of $\lambda$, therefore there is no self-dual invariant Lagrangian of type $L^{\lambda}_{inv}$ for these theories. On the other hand, $L^{\gamma}_{inv}$ has non-vanishing contribution only for the ModMax theory, while the derivative of the Lagrangian density of Maxwell theory with respect to $\gamma$ vanishes. We ensue the following invariant action from the above discussion for the ModMax theory
\begin{eqnarray}\label{LinvMM}
\frac{\partial{\cal L}_{MM}}{\partial\gamma}={\cal L}^{\gamma}_{inv-MM} =\sinh(\gamma) \mathcal{S}+\cosh(\gamma) \sqrt{ \mathcal{S}^2+\mathcal{P}^2}.
\end{eqnarray}

Comparing the self-dual Lagrangian \reef{LinvMM} with Eq. \reef{NN}, we can find a  manifestly $SO(2)$ duality-symmetric form for the Lagrangian \reef{LinvMM} as follows
\begin{eqnarray}\label{NLinMM}
{\cal L}^{\gamma}_{inv-MM}= \frac{1}{8\sinh(\gamma)} \,\, {\cal N}_{\rho}{}^{\rho}.
\end{eqnarray}
It is noticed that there is a similar consideration for the GBI theory \reef{GBI} as well. In fact, variations of the GBI Lagrangian with respect to the  couplings $\lambda$ and $\gamma$, which are respectively denoted by ${\cal L}^{\lambda}_{inv-BI\gamma}$ and ${\cal L}^{\gamma}_{inv-BI\gamma}$, are invariant under $SO(2)$ duality group. In the rest of this section, we will derived the frameworks of these invariant theories in the context of $SL(2,R)$ structures.
\subsection{Self-dual invariant Lagrangian: ${\cal L}^{\lambda}_{inv-BI\gamma}$}
For the purpose of constructing a self-dual Lagrangian for GBI theory, let us briefly review the construction of this duality for an arbitrary electrodynamic theory ${\cal L}(\mathcal{S},\mathcal{P})$ which satisfies the duality invariant condition. Now, one can define an invariant Lagrangian ${\cal L}^{\lambda}_{inv}$ so that is self-dual as follows  \cite{Aschieri:2008ns,Aschieri:2013nda,Chemissany:2006qd,Carrasco:2011jv,Chemissany:2011yv}
\begin{eqnarray}\label{Linv}
 -\lambda\frac{\partial{\cal L}}{\partial\lambda}={\cal L}^{\lambda}_{inv}={\cal L}(\mathcal{S},\mathcal{P})+\frac{1}{4} F_{\mu\nu} G^{\mu\nu}.
\end{eqnarray}

As we mentioned above, due to the lack of dependence on $\lambda$, the invariant actions for both the Maxwell theory and the ModMax theory are trivial. For example, in the case of ModMax theory one can check this fact by inserting ${\cal L}_{MM}$ from \reef{LMM} and $ G^{\mu\nu}$ from \reef{G} in Eq. \reef{Linv}.
  Now, consider a family of  NED theories that led to the Maxwell theory (or the ModMax theory) at the first order of coupling constant, i.e. ${\cal O} ( F^2)$.
  It  seems that the  nonzero self-dual contribution of these NED theories appears from the order of ${\cal O} (F^4)$\cite{Carrasco:2011jv}, but we will show that this statement is not true for the ModMax theories.
By comparing \reef{EMT1} and \reef{Linv}, one can find that ${\cal L}^{\lambda}_{inv}= \frac{1}{4}  {T_{\mu}}^{\mu}$. This interesting result asserts that  the necessary condition for a NED theory to have nonzero contribution to self-dual Lagrangian \reef{Linv} is that ${T_{\mu}}^{\mu}\neq 0$.

The axion-dilaton $BI\gamma$ theory satisfies both duality invariant condition $G\tilde{G}-F\tilde{F}=0$ and the energy-momentum traceless condition ${T_{\mu}}^{\mu}\neq 0$. If we set $\phi_0 = 0$ and $C_0 =0$ in \reef{GBIAD} then we will find the corresponding duality invariant Lagrangian as follows
\begin{eqnarray}\label{GBinv}
{\cal L}^{\lambda}_{inv-BI\gamma}=  \frac{1}{\lambda} \Big[ 1 -  \frac{1}{\sqrt{1 -  \lambda \bigl( 2 {\cal L}_{MM}+\lambda \mathcal{P}^2 \bigr)}} +\lambda \frac{ {\cal L}_{MM}}{\sqrt{1 -  \lambda \bigl( 2 {\cal L}_{MM}+\lambda \mathcal{P}^2 \bigr)}}\Big].
\end{eqnarray}
In other words, a comparison between  Eq.~\reef{Tlanda} for $\phi_0 = 0$ and Eq.~\reef{GBinv} shows that the identity ${\cal L}^{\lambda}_{inv}= \frac{1}{4}  {T_{\mu}}^{\mu}$ works truly. Also, it is worth mentioning that  the $\lambda$-expansion of Lagrangian \reef{GBinv} starts from the order of $\lambda$, or equivalently  ${\cal O}( F^4)$.

Considering the duality invariant structure \reef{Nmunu} in which $G_{\mu\nu}$ is substituted from \reef{Gbar} with $\phi_0= 0$, we find the $\lambda$-expansion of action \reef{GBinv} in the form that is manifestly duality invariant
\begin{eqnarray}\label{Linvlambda}
{\cal L}^{\lambda}_{inv-BI\gamma}= - \frac{\lambda}{2^7 \sinh^2(\gamma)}  {\cal N}_{\mu}{}^{\mu} \, {\cal N}_{\nu}{}^{\nu}+ \frac{\lambda^2 \cosh(\gamma)}{ 2^9 \sinh^4(\gamma)}{\cal N}_{\mu}{}^{\mu} \, {\cal N}_{\nu}{}^{\nu}\, {\cal N}_{\alpha}{}^{\alpha}+\ldots ,
\end{eqnarray}
in which we have used the following identity
\begin{eqnarray}
{\cal N}_{\mu \nu} \, {\cal N}^{\mu \nu} =  \frac{\cosh(2 \gamma)}{4 \sinh^2(\gamma)}  {\cal N}_{\mu}{}^{\mu} \, {\cal N}_{\nu}{}^{\nu}.
\end{eqnarray}

As alluded before, the $SO(2)$-invariant action ${\cal L}^{\lambda}_{inv-BI\gamma}$ can be related to the action \reef{GBI} by a differential flow equation $ -\lambda\frac{\partial{\cal L}_{BI\gamma}}{\partial\lambda}={\cal L}^{\lambda}_{inv}$. Therefore, one can find the relationship of the duality invariant actions \reef{GBinv} to the  irrelevant $T \bar{T}$-like deformation of the GBI theory\cite{Babaei-Aghbolagh:2022uij,Ferko:2022iru,Conti:2022egv} as well as Born-Infeld theory \cite{Conti:2018jho,Ferko:2019oyv,Babaei-Aghbolagh:2020kjg} (the GBI theory at the limit of $\gamma \to 0$), that is ${\cal L}^{\lambda}_{inv}= - \frac18  \lambda O^{\lambda }_{T^2} $ where $O^{\lambda }_{T^2}=T_{\mu\nu}T^{\mu\nu}- \frac{1}{2} {T_{\mu}}^{\mu} {T_{\nu}}^{\nu}$ is an irrelevant $T\bar{T}$  operator.
\subsection{Self-dual invariant Lagrangian: ${\cal L}^{\gamma}_{inv-BI\gamma}$ }
In this subsection, we consider the second  self-dual invariant action coming from the derivative of GBI Lagrangian \reef{GBI} with respect to the parameter $\gamma$. One can find that
\begin{eqnarray}\label{Linvgamma}
{\cal L}^{\gamma}_{inv-BI\gamma}=\frac{\partial{\cal L}_{BI\gamma}}{\partial\gamma}=\frac{ \sinh(\gamma)\mathcal{S}+\cosh(\gamma) \sqrt{ \mathcal{S}^2+\mathcal{P}^2} }{\sqrt{1 -  \lambda \bigl( 2 {\cal L}_{MM}+\lambda \mathcal{P}^2 \bigr)}},
\end{eqnarray}
  therefore, expanding  the invariant action \reef{Linvgamma} up to order $\lambda$ yields that
\begin{eqnarray}\label{LinvGa}
{\cal L}^{\gamma}_{inv-BI\gamma} &\!\!\!=\!\!\!& \sinh(\gamma) \mathcal{S}+\cosh(\gamma) \sqrt{ \mathcal{S}^2+\mathcal{P}^2}  \\
&\!\!\!+\!\!\!& \lambda \biggl(\cosh(2 \gamma)\mathcal{S} \sqrt{ \mathcal{S}^2+\mathcal{P}^2} + \cosh(\gamma)  \sinh(\gamma)(\mathcal{P}^2 + 2  \mathcal{S}^2)\biggr) +{\cal O}( \lambda^2)+\ldots \,\,.\nonumber
\end{eqnarray}
As the same steps that we did in the previous subsection, by considering the invariant structure \reef{Nmunu} and $\phi_0= 0$, we find the above action in the form that is manifestly duality invariant as follows
\begin{eqnarray}\label{NLin}
{\cal L}^{\gamma}_{inv-BI\gamma}= \frac{1}{8\sinh(\gamma)} \,\, {\cal N}_{\mu}{}^{\mu} - \lambda \,\,\frac{1}{64} \frac{\cosh(\gamma)}{\sinh^3( \gamma)}\,\,  {\cal N}_{\mu}{}^{\mu} \, {\cal N}_{\nu}{}^{\nu} +{\cal O}( \lambda^2)+\ldots \,\,.
\end{eqnarray}

 Note that, the  self-dual invariant Lagrangian ${\cal L}^{\lambda}_{inv-BI\gamma} \reef{Linvlambda}$ starts from the order of $F^4$, but the self-dual invariant Lagrangian ${\cal L}^{\gamma}_{inv-BI\gamma}$ \reef{NLin} starts from  the order of $F^2$. Actually, the nonzero contribution of the invariant action ${\cal L}^{\gamma}_{inv-BI\gamma}$ at the Maxwell order can be interested because the other known NED theories have no contribution for their corresponding invariant actions at this order.

In general, by replacing ${\cal N}$ with $\hat{\cal N}$, the results that we have found in this section extend to the case of $\phi_0 \neq 0$ and $ C_0 \neq 0$. Doing so, we extend the $SO(2)$ duality invariant actions \reef{Linvlambda} and \reef{NLin}  to corresponding $SL(2,R)$-invariant cases.
It would be also of interest to notice that the self-dual invariant Lagrangian \reef{Linvgamma} can be related to a marginal  $T \bar{T}$-like deformation of the ModMax theory \cite{Babaei-Aghbolagh:2022uij}. One can show that ${\cal L}^{\gamma}_{inv-BI\gamma}= \frac12  O^{\gamma }_{T^2}$, where $O^{\gamma }_{T^2}=\sqrt{T_{\mu\nu}T^{\mu\nu}- \frac{1}{4} {T_{\mu}}^{\mu} {T_{\nu}}^{\nu}}$ is a marginal operator. It has been discussed in Refs. \cite{Conti:2022egv,Babaei-Aghbolagh:2022kfz,Ferko:2206jsw} that a dimensional reduction of the ModMax theory to two spacetime dimensions corresponds to the continuous marginal root-$T \bar{T}$ deformation of free bosons .
\section{Discussions}\label{4}
 In this paper, we studied the self-dual invariant structures of the conformal nonlinear modification of the Maxwell electrodynamics (ModMax theory) and its generalization. We found an $SL(2,R)$ invariant form for the energy-momentum tensor of the ModMax theory in terms of some $SL(2,R)$ duality-symmetric structures. We have also obtained the manifestly duality invariant structures for the energy-momentum  tensor of the GBI theory.
 
 In Sec.~\ref{secLinv}, we have proposed two actions which are invariant under $SL(2,R)$ symmetry, and consistent with two  irrelevant and marginal  $T\bar{T}$-like deformations. The first self-dual action \reef{GBinv} is related to the standard form of the expansion of the  GBI Lagrangian  with respect to $\lambda$, and the second one \reef{Linvgamma} concerns with the  expansion of the  GBI Lagrangian  with respect to $\gamma$, proposed in Ref.~\cite{Bandos:2020hgy}, which are manifestly $SL(2,R)$-duality invariant and include axion-dilaton couplings.
 
 As custom, the correlation functions are fundamental observables in QFTs. For instance,  it has been shown in Refs.~\cite{He:2020udl,He:2020cxp,He:2021bhj,Song He:2202oyv} that the deformed Lagrangian and stress tensor could obtained order by order from a $T \bar{T}$-product operator in two dimensional CFTs, which can be used to expand the partition function up to the second-order. It would be of interest to study the correlation functions in the context of the  ModMax theory and check out the consistency of  two self-dual actions and   $T \bar{T}$-like deformations.

 The extension of supersyemmetric theories to NED theories are also of great interest \cite{DP,Cecotti:1986gb}. For example, the supersymmetric extension of the Born-Infeld Lagrangian has been studied in Ref.~\cite{Bagger:1996wp} and in the case of ModMax theory this extension was discussed in Refs. \cite{Bandos:2021rqy,Kuzenko:2021cvx}.  Recently, I. Bandos et al. \cite{Bandos:2021rqy} (see also \cite{Kuzenko:2021cvx}) have shown that the Born-Infeld-like extension of  superModMax theory is described by the following superfield Lagrangian density
\begin{eqnarray}\label{susy-BI-MM}
\mathcal{L}_{{ susy}-{ BI \gamma}}
=
\frac{\cosh(\gamma)}{4}\left\{\int
d^2 \theta\, W^2 + \int d^2\bar{\theta} \,\bar{W}^2+\int d^2\theta d^2\bar{\theta}\, W^2 \bar{W}^2K(\mathbb{S},\mathbb{P})\right\}
~,~~~~~~
\end{eqnarray}
where $ K(\mathbb{S},\mathbb{P})$
is given by
\begin{eqnarray}
K(\mathbb{S},\mathbb{P}) =
 \frac{\frac{1}{\lambda} \big(1-\sqrt{1-2\lambda\left[\cosh(\gamma)\mathbb{S}+\sinh(\gamma)\sqrt{\mathbb{S}^2+\mathbb{P}^2}\right]- \lambda^2 \mathbb{P}^2}\big)-\cosh(\gamma)\mathbb{S}}{\cosh(\gamma)(\mathbb{S}^2+\mathbb{P}^2)}
~.~~~~~~
\end{eqnarray}
 Here, the superfields $\mathbb{S}$ and $\mathbb{P}$ are
\begin{eqnarray}
\mathbb{S}
=
-\frac{1}{16}(D^2 W^2+\bar{{D}}^2\bar{W}^2)~,
\quad \mathbb{P}
= \frac{i}{16}({D}^2 W^2-\bar{{D}}^2 \bar{W}^2)
~,
\label{superSandP}
\end{eqnarray}
where $W^2=W^{\alpha} W_{\alpha}$. $W_{\alpha}$ and its conjugate $\bar{W}_{\dot{\alpha}}=(W_\alpha)^*$, are anticommuting Weyl spinor chiral superfields satisfying in the conditions
\begin{gather}
\bar{{D}}_{\dot{\beta}}W_{\alpha} = 0, \quad {D}^{\alpha}W_{\alpha}=\bar{{D}}_{\dot{\alpha}}\bar{W}^{\dot{\alpha}} ~.
\end{gather}

In the limit  $\gamma=0$, the Lagrangian \eqref{susy-BI-MM} reduces to the Bagger-Galperin Lagrangian \cite{Bagger:1996wp}. The self-dual condition for electromagnetic duality invariance of generic nonlinear electrodynamic theories in Refs. \cite{Gaillard:1981rj,Gaillard:1997rt}, were generalized to superfield formulations of $N = 1$  supersymmetric theories in Refs.\cite{Kuzenko:2000tg,Kuzenko:2000uh}. For generic $N = 1$ theories described by a Lagrangian
$\mathcal{L}[W, \bar{W} ]$, the Kuzenko–Theisen duality-invariance condition is \cite{Kuzenko:2000tg,Kuzenko:2000uh}
\begin{eqnarray}\label{SDC}
Im\int d^4x\,d\theta^2\left( W^\alpha W_\alpha+M^\alpha M_\alpha \right)=0\, , \qquad
M_\alpha\equiv-2i\frac{\delta \mathcal{L}[W,\bar W]}{\delta W^\alpha} \, .
\end{eqnarray}
The  Born-Infeld-like extension of the supersymmetric ModMax theory in \reef{susy-BI-MM} satisfies the duality-invariance condition \reef{SDC}. Therefore, as shown in Sec. \ref{secLinv},  we expect to have two supersymmetric self-duality invariant actions for the $ N = 1 $ supersymmetric Born-Infeld-like theory.
These two actions are denoted by
\begin{eqnarray}
 \mathcal L_{{ susy}-inv- BI \gamma}^\lambda=- \lambda \frac{ \partial \mathcal L_{{ susy}- BI \gamma}}{\partial \lambda}
\,\,\,\,\,\,\,\,;\,\,\,\,\,\,\, \mathcal L_{{ susy}-inv- BI \gamma}^\gamma=\frac{ \partial \mathcal L_{{ susy}- BI \gamma}}{\partial \gamma}
~.
\label{flow-susy-BIMM}
\end{eqnarray}

 In \cite{Babaei-Aghbolagh:2020kjg} we have, a priori, studied the compatibility deformation of $N=2$ supersymmetric BI theory with $N=2$ supersymmetric self-dual invariant Lagrangian.
The $T\bar{T}$-deformation of the $N=1$ supersymmetric ModMax theory with operator $O_{T^2}^{\lambda}$, was initially considered  in Ref. ~\cite{Ferko:2022iru}.
 As final remark, extending the $N=1$ and $N=2$ supersymmetric self-dual invariant Lagrangian of the ModMax theory  can be investigated as a future work.

\section*{Acknowledgements}
We are very grateful to D. Sorokin for his kindly interest in this work and fruitful discussion. HBA is specially appreciate M.R.Garousi for his useful comments at the early stages of the work.
\section*{Appendix}\label{App}
In this appendix we write the explicit forms of invariant structures  $\hat{\cal N}_{\rho}{}^{\rho}$
 and $\hat{\cal N}_{\mu}{}^{\rho}\, \hat{\cal N}_{\nu \rho}$ that appear in Eq \reef{hatT}.
\begin{eqnarray}\label{Nhat}
\hat{\cal N}_{\rho}{}^{\rho}&=& 8\, e^{-\phi_0} \sinh(\gamma) \Bigl(\cosh(\gamma)  \sqrt{ \mathcal{S}^2+\mathcal{P}^2} + \sinh(\gamma)\mathcal{S}\Bigr)  \\
&+&  \lambda \, e^{-2\phi_0} \Bigl[4 \cosh(\gamma) \mathcal{P}^2 + 4 \cosh(3 \gamma) (\mathcal{P}^2 + 2  \mathcal{S}^2) + 8 \sinh(3 \gamma)\mathcal{S} \sqrt{ \mathcal{S}^2+\mathcal{P}^2}\Bigr]+{\cal O} ( \lambda^2)+\ldots\, ,\nonumber
\end{eqnarray}
and
\begin{eqnarray}
\hat{\cal N}_{\mu}{}^{\rho}\, \hat{\cal N}_{\nu \rho} &=&- e^{-2\phi_0}  \frac{2 \sinh(2 \gamma) \bigl(\mathcal{P}^2 + \cosh(2 \gamma) (\mathcal{P}^2 + 2  \mathcal{S}^2) + 2 \sinh(2 \gamma)\mathcal{S} \sqrt{ \mathcal{S}^2+\mathcal{P}^2}\bigr)}{\sqrt{ \mathcal{S}^2+\mathcal{P}^2}}  T^{Max}_{\mu \nu} \\
&+& 2 \, e^{-2\phi_0} \cosh(2 \gamma) \Bigl(\mathcal{P}^2 + \cosh(2 \gamma) (\mathcal{P}^2 + 2  \mathcal{S}^2) + 2 \sinh(2 \gamma)\mathcal{S} \sqrt{ \mathcal{S}^2+\mathcal{P}^2}\Bigr) g_{\mu \nu}  \nonumber \\
&+& \lambda \, e^{-3\phi_0}  \biggl[\Bigl(-2 \cosh(\gamma) \mathcal{P}^2 - 4 \cosh(3 \gamma) \mathcal{P}^2 - 2 \cosh(5 \gamma) (\mathcal{P}^2 + 4  \mathcal{S}^2) \nonumber \\
&+&  \frac{32 \cosh^2(\gamma)  \sinh(\gamma)\mathcal{P}^2\mathcal{S}}{\sqrt{ \mathcal{S}^2+\mathcal{P}^2}} -  \frac{24 \cosh^2(\gamma)  \sinh(3 \gamma)\mathcal{P}^2\mathcal{S}}{\sqrt{ \mathcal{S}^2+\mathcal{P}^2}} -  \frac{8  \sinh(5 \gamma)\mathcal{S}^3}{\sqrt{ \mathcal{S}^2+\mathcal{P}^2}}\Bigr) T^{Max}_{\mu \nu}  \nonumber \\
&+& \Bigl(-2 \cosh(\gamma) \mathcal{P}^2 \mathcal{S} + 4 \cosh(3 \gamma) \mathcal{P}^2\mathcal{S}+ \cosh(5 \gamma) (6 \mathcal{P}^2\mathcal{S}+ 8 \mathcal{S}^3) +  \sinh(\gamma)2 \mathcal{P}^2 \sqrt{ \mathcal{S}^2+\mathcal{P}^2}\nonumber \\
&+& \bigl(4 \mathcal{P}^2 \sinh(3 \gamma) + 2 \sinh(5 \gamma)(\mathcal{P}^2 + 4  \mathcal{S}^2)\bigr)\Bigr) \mathit{g}_{\mu \nu} \sqrt{ \mathcal{S}^2+\mathcal{P}^2}\biggr]+{\cal O} ( \lambda^2)+\ldots\, .\nonumber
\end{eqnarray}

\if{}
\bibliographystyle{abe}
\bibliography{references}{}
\fi

\providecommand{\href}[2]{#2}\begingroup\raggedright\endgroup
\end{document}